\documentclass[article,superscriptaddress,amsmath,amssymb,aps,onecolumn,nofootinbib]{revtex4-1}
\usepackage{tabularx}
\usepackage{graphicx}
\usepackage{euscript}
\usepackage{epsfig,psfrag,subfigure}
\usepackage{dcolumn}
\usepackage{bm}
\usepackage{setspace}
\usepackage{floatrow}
\usepackage{appendix}
\usepackage{color}
\usepackage{amsfonts}
\usepackage{exscale}
\usepackage{multirow}
\usepackage{hyperref}
\hypersetup{colorlinks=false,linkcolor=black}
\usepackage[left]{lineno}
\usepackage{comment}
\usepackage{ulem}
\usepackage{braket}
\usepackage{bbm}
\def\be{\begin{equation}}       \def\ee{\end{equation}}
\def\bea{\begin{eqnarray}}      \def\eea{\end{eqnarray}}

\newcommand{\PreserveBackslash}[1]{\let\temp=\\#1\let\\=\temp}
\newcolumntype{C}[1]{>{\PreserveBackslash\centering}p{#1}}
\newcommand{\argmax}{\mathop{\mathrm{arg\,max}}}

\begin{document}
\captionsetup[figure]{labelfont={bf}}

\title{A quantum-inspired tensor network method for constrained combinatorial optimization problems}

\author{Tianyi Hao}
\affiliation{Department of Computer Science, Stanford University, Stanford, California 94305, USA}

\author{Xuxin Huang}
\affiliation{Department of Applied Physics, Stanford University, Stanford, California 94305, USA}
\affiliation{Stanford Institute for Materials and Energy Sciences, SLAC National Accelerator Laboratory, Menlo Park, California 94025, USA}



\author{Chunjing Jia}
\email{chunjing@stanford.edu}
\affiliation{Stanford Institute for Materials and Energy Sciences, SLAC National Accelerator Laboratory, Menlo Park, California 94025, USA}
\affiliation{Department of Physics, University of Florida, Gainesville, Florida  32611, USA}

\author{Cheng Peng}
\email{cpeng18@stanford.edu}
\affiliation{Stanford Institute for Materials and Energy Sciences, SLAC National Accelerator Laboratory, Menlo Park, California 94025, USA}

\begin{abstract}
Combinatorial optimization is of general interest for both theoretical study and real-world applications. Fast-developing quantum algorithms provide a different perspective on solving combinatorial optimization problems. In this paper, we propose a quantum-inspired tensor-network-based algorithm for general locally constrained combinatorial optimization problems. Our algorithm constructs a Hamiltonian for the problem of interest, effectively mapping it to a quantum problem, then encodes the constraints directly into a tensor network state and solves the optimal solution by evolving the system to the ground state of the Hamiltonian. We demonstrate our algorithm with the open-pit mining problem, which results in a quadratic asymptotic time complexity. Our numerical results show the effectiveness of this construction and potential applications in further studies for general combinatorial optimization problems.
\end{abstract}
\maketitle
\tableofcontents
\section{Introduction}
Combinatorial optimization is the process of finding an optimal object from a discrete and finite set of objects. Combinatorial optimization has extensive applications in almost every field of industry, such as supply chain optimization \cite{supplychain}, transportation networks and power grids \cite{powergrids}, and finance \cite{cobook}. The search space of a combinatorial optimization problem increases rapidly with the problem size. Problems like the Boolean satisfiability problem can have an exponentially large solution space, making an exhaustive search inapplicable for large-scale problems. From the complexity theory perspective, many combinatorial problems fall into the class of {$\textsf{NP-hard}$}, which is generally believed to be unsolvable in polynomial time on classical computers. Classical algorithms often employ heuristics and approximations to find nearly optimal solutions \cite{lin1973effective}. Quantum algorithms \cite{montanaro2016quantum}, on the other hand, harness the power of randomness, superposition, entanglement, and interference from quantum mechanics, { which might lead to an advantage in exploring the solutions of a combinatorial optimization problem} \cite{moll2018quantum, QAOA2014arxiv, VQEnatcommun}. The implementation of quantum algorithms is currently limited by small-scale, noisy, and error-prone contemporary hardware \cite{preskill2018quantum}; nevertheless, they view the problems from a different perspective, motivating many quantum-inspired classical algorithms to appear \cite{han2002quantum, tang2019quantum}.

Tensor networks (TNs) have undergone rapid development in the last two decades, gaining tremendous success in quantum many-body physics, quantum information sciences, statistical physics, and so on. Tensor network algorithms based on matrix product states (MPS) \cite{mpsmath1,mpsmath2}, projected entangled pair states (PEPS) \cite{VerstraetePRA,VerstraeteARXIV}, and variational renormalization group methods \cite{NRG2975wilson,dmrg1992white,dmrg1993white} are very efficient in simulating a large class of quantum many-body systems. The tensor network structure can encode the combinatorial optimization problem with local constraints, providing an idea of utilizing the tensor network to solve combinatorial optimization problems. 

This paper presents a quantum-inspired tensor network algorithm to solve constrained combinatorial optimization problems and demonstrates the algorithm with a particular problem with numerical results. The paper is structured as follows: The general quantum-inspired tensor network algorithm is proposed in Section 2, and the open-pit mining problem is provided as an example in Section 3. Section 4 shows our numerical results with the quantum-inspired tensor network algorithm and concludes with a discussion of open questions and directions for future work.



\section{A general tensor network algorithm for combinatorial optimization}
In this section, we divide our algorithm into four components and describe each in detail. Section \ref{sec:problem_mapping} explains how to construct the Hamiltonian for a classical combinatorial problem to transform it to a quantum problem. Section \ref{sec:postselection} serves as inspiration for our core idea, Section \ref{sec:tnconstruction}, where the former studies how to satisfy the constraints without introducing a penalty term, and the latter details how to construct a tensor network state that represents the superposition of all feasible solutions. Finally, in Section \ref{ITE}, we show how to find the optimal solution by evolving the tensor network state to the ground state of the problem Hamiltonian.

\subsection{Hamiltonian construction/Problem mapping}\label{sec:problem_mapping}
    We use an objective function $f$ and a discrete set of feasible solutions $\textbf{x}$ to specify the combinatorial optimization problem. In many such problems, each solution involves a binary selection of the individual components under certain constraints, denoted as $\textbf{x}=(x_1,\cdots,x_n)$ with binary variables $x_n\in \{0,1\}$, satisfying constraints $\textbf{c}=(c_1,\cdots,c_m)$. The goal is to find the solution $\textbf{x}^\star$ in all feasible solutions to maximize the objective function $f$, {\it i.e.} $\textbf{x}^\star=\argmax_{\textbf{x}\in \{0,1\}^{\otimes n}} f(\textbf{x})$. This goal can be achieved by mapping the problem to a Hamiltonian and looking for the ground state in the Hilbert space. The Hamiltonian is written as
    \begin{align}
        \hat{H}=\sum_{\mathbf{x}\in\{0,1\}^{\otimes n}}a_{\mathbf{x}}\ket{\mathbf{x}}\bra{\mathbf{x}},
    \end{align}
    where $a_{\textbf{x}}$ are real values representing the elements of the Hamiltonian and $|\textbf{x}\rangle=|x_1\rangle\otimes|x_2\rangle\cdots\otimes|x_n\rangle$. Note that in this paper, we are interested in problems that require $n$ qubits, where $n$ is the number of the binary variables. Each variable $x_i$ of the set of solutions $\textbf{x}$ is assigned on the basis of the Pauli operator $\hat{Z}_i$ on the $i$-th qubit, that is, $\hat{Z}_i = \ket{0}\bra{0} -\ket{1}\bra{1}$ with $\ket{x_i}\in\{\ket{0},\ket{1}\}$. Thus, a general quantum state can be expressed as
    \begin{align}
    \ket{\Psi}=\sum_{\mathbf{x}\in\{0,1\}^{\otimes n}} b_\mathbf{x}\ket{\mathbf{x}}.
    \end{align}
    Here $b_\mathbf{x}$ represents the linear coefficient that meets the normalized condition, {\it i.e.} $\sum_{i=1}^n |b_{x_i}|^2=1$. Consider a general unconstrained binary linear optimization problem:
    \begin{align}
        \min\{\mathbf{w}^T\mathbf{x}: \mathbf{x}\in \{0,1\}^{\otimes n}\}
    \end{align}
    with weights $\mathbf{w}=(w_1,\cdots,w_n)$. Then the Hamiltonian is transformed as
    \begin{align}
        \hat{H} &= \sum_{i} \frac{w_i}{2}(\hat{Z}_i-\hat{I}),\label{Eq:Hamilton}
    \end{align}
    where $I$ is the identity. 
    Such transformations take linear time in the number of variables.\\
    The ground state of this Hamiltonian, denoted as $\ket{\psi_g}$, 
    is the state that minimizes the energy defined as $E(\ket{\Psi})=\bra{\Psi}\hat{H}\ket{\Psi}/\langle \Psi\ket{\Psi}$. 
    With an appropriate mapping, the original optimization problem is equivalent to solving the ground state of the Hamiltonian. Then the ground-state solving techniques in quantum many-body physics can be utilized to achieve an optimized solution of the combinational optimization problem.

\subsection{Postselection: a penalty-free approach}\label{sec:postselection}
    For constrained problems, additional terms are usually needed in the Hamiltonian to penalize constraint violations. These penalty terms should be conditioned by multiplying with appropriate penalty factors to ensure that the overall Hamiltonian behaves as intended. Optimization of such a Hamiltonian containing penalty terms inevitably involves tuning of additional hyperparameters, which greatly increases the difficulty of the overall optimization process. 
    
    In our study, we present a hyperparameter-free algorithm that directly projects a randomized initial quantum state into a subspace that satisfies the constraints. 
    We use a projector, defined as $\hat{P}\ket{\Psi} = |\Psi_{v}\rangle$, to eliminate states that violate constraints. $\ket{\Psi}$ is an arbitrary state in the Hilbert space, and $|\Psi_v\rangle$ is the projected state that belongs to the subspace denoted by $V$ satisfying the constraints. The projector is then constructed as
    \begin{align}
        \hat{P}=\sum_{\mathbf{x}\in V}|\mathbf{x}\rangle\langle \mathbf{x}|.
    \end{align}
   It satisfies $\hat{P}^{2}=\hat{P}$. The ground state $\ket{\psi_g}$ under the projection is written as follows:
    \begin{align}
        \ket{\psi_g}&=\min_{\ket{\Psi}}\ E(\ket{\Psi}),\\
        E(\ket{\Psi})&=\frac{\bra{\Psi_{v}}\hat{H}\ket{\Psi_{v}}}{\braket{\Psi_{v}|\Psi_{v}}}=\frac{\bra{\Psi}\hat{P}\hat{H}\hat{P}\ket{\Psi}}{\braket{\Psi|\hat{P}|\Psi}}.
    \end{align}
    We can then extract the optimal solution to the combinatorial optimization
    problem encoded in $\ket{\psi_g}$. 
    It is worth noting that the solver only works for a normalized state, {\it i.e.} $\langle\Psi\ket{\Psi}=1$, while the projected state are not normalized since a general projector $\hat{P}$ is not unitary. However, with prior knowledge that the ground state is always a product state following the constraints, and thus $\braket{\Psi_0|\hat{P}|\Psi_0}=1$, we could simply apply the solver for the normalized ground state of $\hat{P}\hat{H}\hat{P}$.
    

\subsection{Tensor network construction of the projected state}\label{sec:tnconstruction}
    
    Directly constructing the projector $\hat{P}$ incurs an exponential cost in time and memory. However, by encoding the constraints in a tensor network state, we can significantly reduce the cost to $O(mn)$, or $O(m+n)$ if the constraints are ``local", {\it i.e.} the number of variables involved in each constraint is limited by a constant and vice versa. 
    The general form of the tensor network state can be written as
    \begin{align}
        \ket{\Psi} = \sum_{\{\alpha, \beta \cdots\}}\prod_{i=1}^{n} T^{[x_i]}_{\alpha_{i} \beta_1\cdots\beta_x } R^{[c_i]}_{\beta_1\cdots\beta_c}|\alpha_{1}\cdots \alpha_{i} \cdots\alpha_{n}\rangle. \label{Eq:generalTN}
    \end{align}
    
    There are two types of tensors, $T^{[x]}$ and $R^{[c]}$. Each $T^{[x]}$ has a single physical index labeled $\alpha$ with a bond dimension $d=2$ reflecting the binary variable, that is, $\alpha=0$ and $1$, and multiple virtual indices labeled $\beta$ also with the same bond dimension. Each auxiliary tensor $R^{[c]}$ encodes the constraint $c$ and connects to $T^{[x]}$ tensors whose variable $x$ is involved in the constraint $c$. 
    To map the selection of the binary variable from the physical index to other indices, $T^{[x]}_{0\cdots 0}$ and $T^{[x]}_{1\cdots 1}$ are set to $1$. The $R^{[c]}$ tensors are constructed by traversing the allowed assignments of the constrained variables and setting the elements located in the corresponding indices to $1$. We initialize the $T^{[x]}$ and $R^{[c]}$ tensors as zeros. For example, if we have variables $x_1, x_2, x_3 \in \{0,1\}$ and a single constraint $c_1: x_1 + x_2 + x_3 \mod 2 = 1$, then we can use four sparse tensors, {\it i.e.} $T^{[x_1]}_{\alpha\beta_1}, T^{[x_2]}_{\alpha\beta_2}, T^{[x_3]}_{\alpha\beta_3}$ and $R^{[c_1]}_{\beta_1\beta_2\beta_3}$ to encode this particular constrained problem, with nonzero elements of the tensors specified as:
    
    \begin{align}
        T^{[x_1]}_{00}=T^{[x_1]}_{11}=1&\\
        T^{[x_2]}_{00}=T^{[x_2]}_{11}=1&\\
        T^{[x_3]}_{00}=T^{[x_3]}_{11}=1&\\
        R^{[c_1]}_{001}=R^{[c_1]}_{010}=R^{[c_1]}_{100}=R^{[c_1]}_{111}=1&.
    \end{align}
    
\begin{figure}[tbh!]
\centering
    \includegraphics[width=0.7\linewidth]{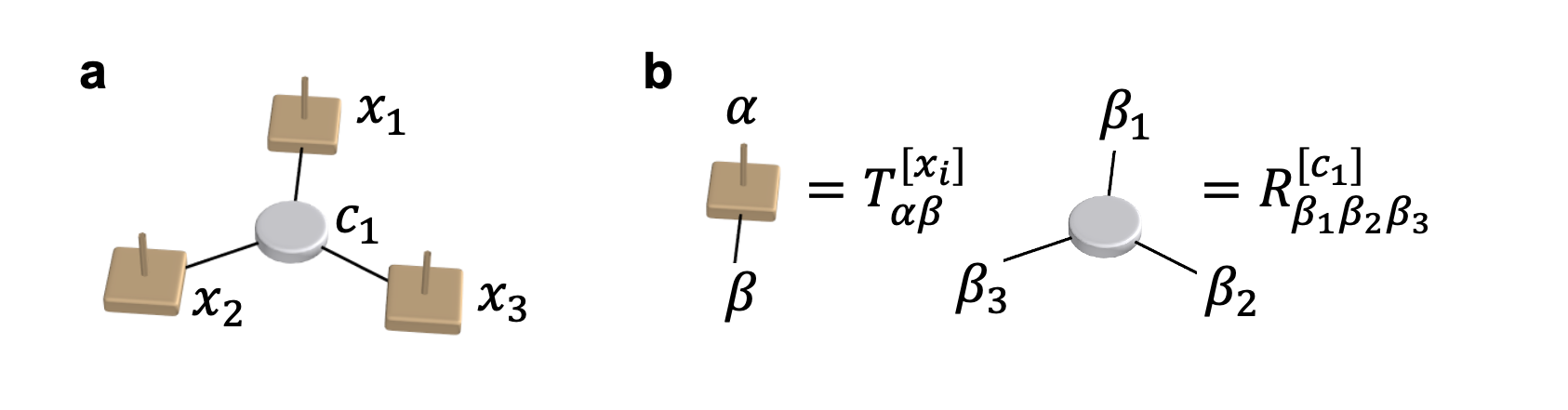}
    \caption{\textbf{Tensor network construction for a three-variable constrained problem.} (a) The schematic for the three-variable constrained problem with $x_1, x_2, x_3 \in \{0,1\}$ and the constraint $c_1$. (b) The tensor definitions in the general form of the tensor network for the three-variable constrained problem, with $\alpha$ as the physical index and $\beta$ as the virtual index.}
\label{Fig:simple_tensor}
\end{figure}

    The four possibilities to satisfy the constraint $c_1$ have been included in $R^{[c_1]}$; for example, $R^{[c_1]}_{001}=1$ with indices $001$ represents $x_1=x_2=0$ and $x_3=1$. The construction of these tensors is visualized in Fig.\ref{Fig:simple_tensor}. 
    
    By this construction, we encode all feasible variable assignments in the initialized tensor network. Since traversing the allowed assignments of each constraint requires exponential time in the number of variables involved, the overhead of our method is much lower for problems with local constraints, compared to directly building the projector. Additionally, existing tensor algebra methods and tensor network algorithms are applicable to structured problems, helping us solve problems efficiently.

\subsection{Finding the optimal solution}\label{ITE}
    The optimal solution is one of the basis vectors in Eq.~(\ref{Eq:generalTN}), denoted by $|\alpha_1 \cdots \alpha_i \cdots\alpha_n\rangle$, with $\alpha_i=0,1$ on the $i$-th binary variable $x_i$. 
    Having the superposition of all allowed variable assignments, we can screen out the optimal solution by imaginary time evolution (ITE), a technique to project the initial state to the ground state of an objective Hamiltonian. 
     ITE effectively performs a power iteration by repetitively applying the ITE operator $e^{-\tau\hat{H}_p}$ to find the ground state of $\hat{H}$. The ground state minimizes the energy, which is equivalent to minimizing or maximizing the objective function. The simulation of the ITE takes the form of 
    \begin{align}\label{Eq:ITE1}
    \ket{\Psi} = \lim_{\tau\rightarrow\infty}\frac{e^{-\tau\hat{H}}\ket{\Psi_{0}}}{||e^{-\tau\hat{H}}\ket{\Psi_{0}}||},
    \end{align}
    where $\tau$ is referred to as the imaginary time. If the Hamiltonian only contains the summation of commuting terms, then $e^{-\tau\hat{H}}$ can be directly rewritten into the product of the local evolution operators, {\it i.e.} $e^{-\tau\hat{H}}=\Pi_i e^{-\tau\hat{h}_i}$. Otherwise, the Trotter-Suzuki decomposition\cite{Suzuki1991,SUZUKI1990319}, an approximation for doing the ITE if containing non-commuting terms in the Hamiltonian, should be involved.
    
    To extract the optimal solution from the tensor networks, we take inspiration from measuring the spin magnetic moment. The assignment of the variable $x_i$ is calculated as
    
    \begin{align}\label{Eq:ITE2}
    x^\star_i = \mathbbm{1}(\frac{\braket{\Psi| \hat{Z}_{i}|\Psi}}{\braket{\Psi|\Psi}} < 0),
    \end{align}
    where $\hat{Z}_{i}$ is the Pauli matrix. That is, if the expectation value is negative, $x_i$ is $1$; otherwise, it is $0$. Since we need to do this calculation for each variable, there is an $O(n)$ factor multiplied with the complexity of computing a single expectation value, which we will discuss shortly.
    
    Note that there could be certain variables whose either assignment gives the same final objective value and obeys the constraints, known as the degeneracy. In this case, if we prefer $x_i$ to be assigned a specific value, for example, zero, then the measurement operator can be adjusted as 
    
    \begin{align}\label{Eq:ITE3}
    \hat{\tilde{Z}}_{i} = \begin{bmatrix} 1 & 0 \\ 0 & -\mu \end{bmatrix},
    \end{align}
    where $\mu$ is a value larger than $1$ so as to create a slight difference between the expectation values of the $0$ and $1$ variable assignments. Then we should measure
    \begin{align}\label{Eq:ITE4}
    \tilde{x}^\star_i = \mathbbm{1}(\frac{\braket{\Psi| \tilde{\hat{Z}}_i|\Psi}}{\braket{\Psi|\Psi}} < \nu),
    \end{align}
    where $\nu$ is a threshold value related to $\mu$ used to detect the slight difference. In practice, $\mu$ can be slightly larger than $1$ and $\nu$ is a small positive number approaching zero, such as $\mu=3$ and $\nu=0.04$, both depending on how many degenerate states exist. We can adjust $\mu$ until the degenerate states are properly separated, then we can find our preferred one by setting an appropriate threshold $\nu$.
    
    We calculate the expectation values $\bra{\Psi} \tilde{\hat{Z}}_{i}\ket{\Psi}$ by performing tensor network contractions \cite{Ran2020, pang2020efficient, guo2019general}, which sometimes take exponential time and space to obtain an exact result. 
    Tensor decomposition methods, such as singular value decomposition (SVD), are regularly used in tensor network contractions to limit the rapidly increasing bond dimension. Specifically, we can preserve a predetermined number of columns and rows in the decomposed matrices with the largest singular values. An alternative common practice is to preserve columns and rows whose singular values are above a threshold. Using these techniques cleverly can reduce the complexity of the whole contraction to be polynomial in $m$ and $n$, while still resulting in an accurate enough value. Moreover, for problems that have a structured tensor network construction, we can utilize existing tensor network algorithms to further optimize performance.

\section{The open-pit mining problem}
\subsection{Problem description}\label{sec:mining_problem}
   We use a real-world combinatorial optimization problem, the open-pit mining problem, as an example to demonstrate our algorithm. The goal of designing optimal open-pit mines is to maximize ore production while avoiding unnecessary mining of rocks. The planning is also subject to a variety of constraints on the size, shape, and form of the mine, making it a computationally taxing process.
  Theoretical studies of the problem often apply simplifying assumptions, converting the problem into a simpler and more well-understood form \cite{lerchs1965optimum}. 
  
\begin{figure}[tbh!]
\centering
    \includegraphics[width=0.85\linewidth]{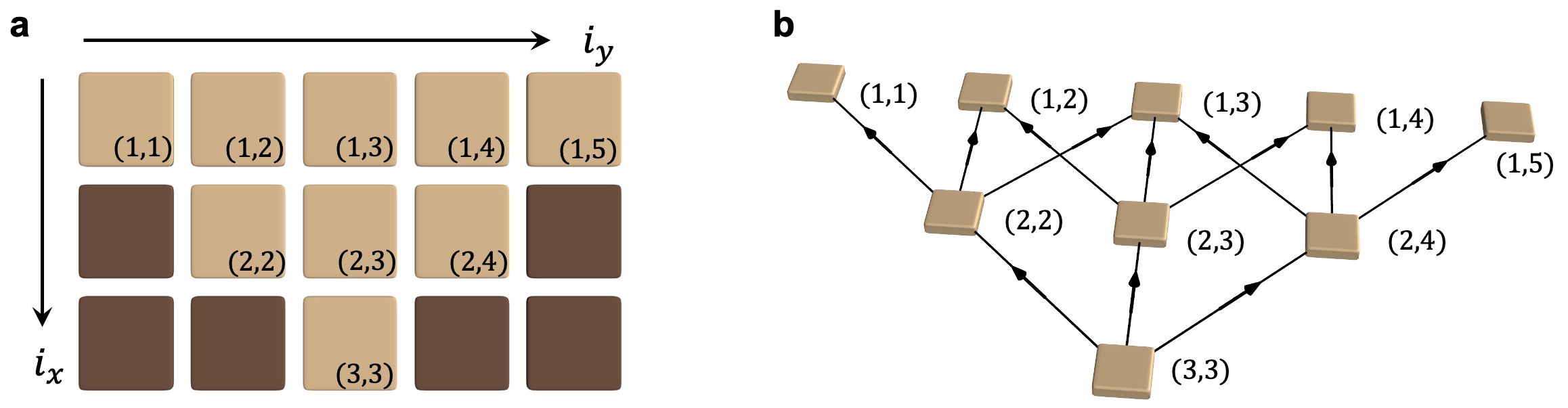}
    \caption{\textbf{Two-dimensional open-pit mining problem and the equivalent graph representation.} (a) A schematic of two-dimensional open-pit mining problem. The blocks in light brown are excavatable. The blocks in dark brown are unexcavatable due to $45^{\circ}$ slope constraint. $i_x$ and $i_y$ represent the vertical and horizontal indices, respectively. (b) The schematic for a directed graph representing a problem equivalent to the two-dimensional open-pit mining in panel (a). Each node $i=(i_x,i_y)$ represents a mining block. Each directed edge $(i, j)\in V$ represents the physical constraint: if block $i$ is excavated, block $j$ should be excavated too. }
\label{Fig:mining_hamiltonian}
\end{figure}

In particular, the 2D open-pit mining problem can be formally stated as a combinatorial optimization problem. Consider a 2D square lattice of mining blocks, where each block $i$ has an associated profit $w_{i}$. The coordinate of block $i$ is denoted as $(i_{x}, i_{y})$. A feasible solution should follow physical constraints, {\it i.e.} if the block $i$ is excavated, then all its child blocks $j$ should be excavated as well. In this work, we consider the $45^{\circ}$ slope constraint: $j\in\{(i_x-1, i_y-1),(i_x-1, i_y),(i_x-1, i_y+1)\}$, as illustrated in Fig.~(\ref{Fig:mining_hamiltonian}) (a). Equivalently, we can consider an $n$-node directed graph $G=(V,E)$ with node weight $w_{i}$, as shown in Fig.~(\ref{Fig:mining_hamiltonian}) (b). $G$ is structured into levels, where each level contains two more nodes than the previous level, starting from one node in the first level. Each node before the last level has exactly three child nodes at the next level, and each node after the first level has one to three parent nodes.

By assigning $x_{i}=0$ (unexcavated) or $x_{i}=1$ (excavated) to each node $i$, we can write the open-pit mining problem as
\begin{align}
    \max \sum_{i=1}^n w_{i}x_{i}
\end{align}
subject to
\begin{align}
    x_i \in \{0,1\} &\text{\ \ \ for } i = 1,2,\cdots,n\\
    \sum_{j\in \mathit{children}(i)} x_{i}(1-x_{j}) = 0 &\text{\ \ \ for } i = 1,2,\cdots,n,
\end{align}
where $\mathit{children}(i)$ denotes the set of child nodes of node $i$.

Traditionally, the open-pit mining problem is solved by reducing to the maximum closure problem or the maximum flow problem and utilizing efficient graph algorithms. In particular, the Lerchs-Gorssman (LG) algorithm \cite{giannini1990optimum,khalokakaie2000lerchs} was the most widely used algorithm in the mining industry, giving a provably optimal solution in polynomial time. In recent years, it has been surpassed by the more efficient Pseudoflow algorithm \cite{hochbaum2008pseudoflow,muir2008pseudoflow}, a $O(|V||E|\log |V|)$ algorithm for the maximum flow problem. 
Recently, a quantum computing approach was proposed as the first attempt to solve this problem with quantum computers \cite{hindy2021quantum}. It modifies the objective function to
\begin{align}
    \max (\sum_{i} w_{i}x_{i} - \lambda \sum_{i, j\in p(i)} x_{i}(1-x_{j})),
\end{align}
where $\lambda$ is a hyperparameter introduced to regularize the penalty for constraint violations. Then the problem Hamiltonian is constructed using the method described in Section \ref{sec:problem_mapping}.

Our algorithm takes inspiration from the quantum computing approach but is intrinsically different. Using tensor networks to represent quantum states, our algorithm can employ powerful non-unitary operations, which are unavailable to quantum algorithms. This fact allows us to directly construct the superposition state of all feasible solutions and avoids having to optimize the regularization of the penalty term. Our algorithm is also completely different from the graph algorithms. It provides a new perspective on such problems, utilizing the rapidly developing tensor network methods. Moreover, the core ideas can be applied to general combinatorial problems, not limited to just this one.

\subsection{The tensor network framework}\label{sec:mining_framework}
\begin{figure}[tbh!]
\centering
    \includegraphics[width=0.85\linewidth]{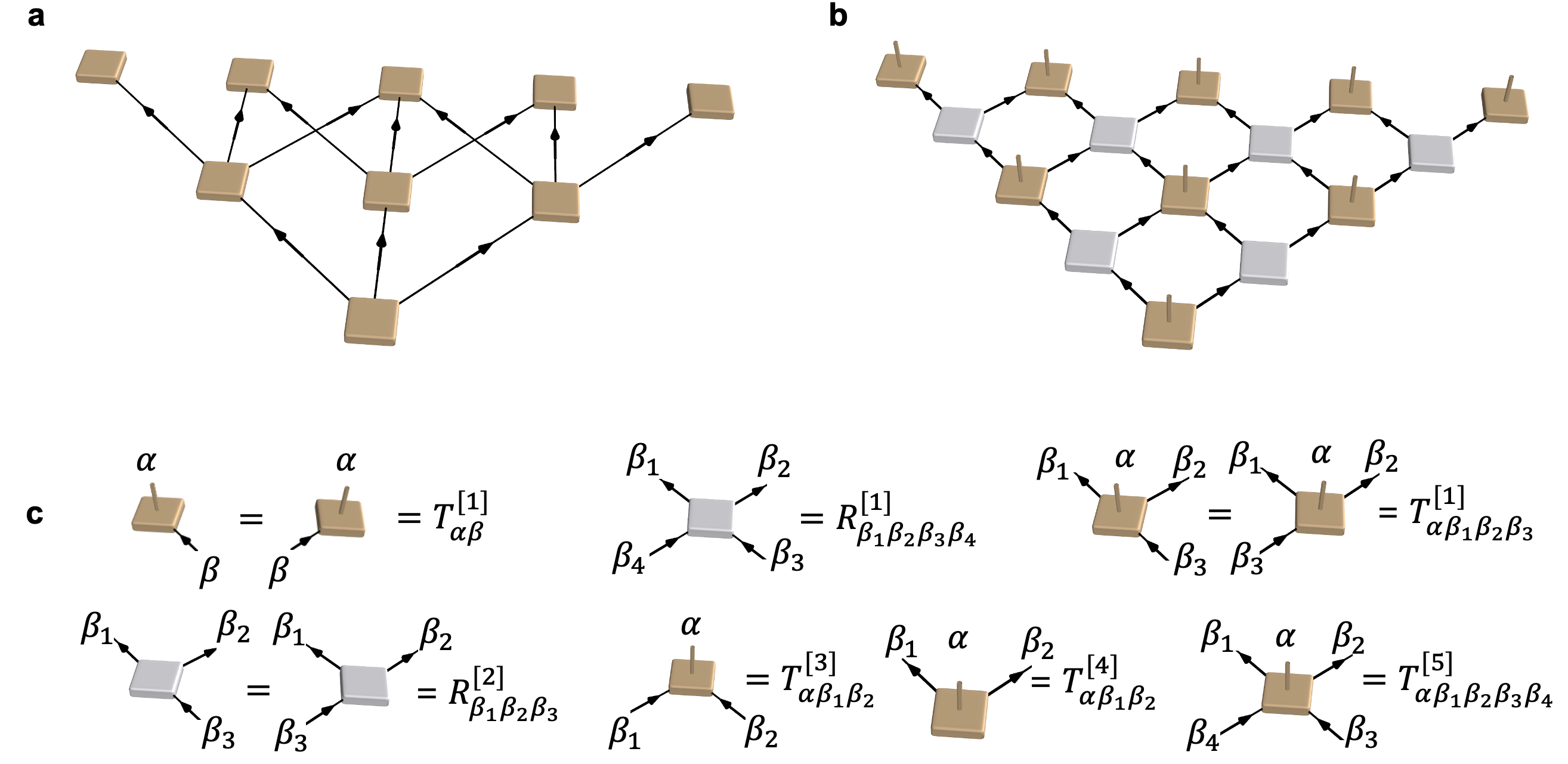}
    \caption{\textbf{Tensor network construction for open-pit mining problem.} (a) Directed graph representation of the open-pit mining problem, with the excavatable blocks expressed as nodes in light brown. (b) A corresponding tensor network construction, where the light brown blocks show the physical nodes and the gray blocks show the virtual nodes. (c) The definition of nonequivalent tensors in our tensor network construction. $\alpha$ and $\beta$ represent the physical and virtual indices, respectively.}
\label{Fig:def_peps_2d}
\end{figure}

We construct the configurations of all allowed states obeying the smoothness constraints as a tensor network state. Fig.\ref{Fig:def_peps_2d} visualizes the construction process using the $5 \times 3$ mine as an example.
Referring to the smoothness constraints that the walls of the pit should not exceed a maximum steepness, the sites marked dark brown in Fig.\ref{Fig:mining_hamiltonian}(a) would, if excavated, violate the smoothness constraints, which therefore should be excluded from the pit profile as well as the initialized tensor network construction. The values of the tensors as defined in Fig.\ref{Fig:def_peps_2d}(c) are as follows (here we group the tensors corresponding to different variables or constraints but have the same form together):
\begin{eqnarray}
&T^{[1]}_{11}=T^{[1]}_{00}=1\\
&T^{[2]}_{1111}=T^{[2]}_{0000}=1\\
&T^{[3]}_{111}=T^{[3]}_{000}=1\\
&T^{[4]}_{111}=T^{[4]}_{000}=1\\
&T^{[5]}_{11111}=T^{[5]}_{00000}=1\label{Eq:t5}\\ 
&R^{[1]}_{1111}=R^{[1]}_{1110}=R^{[1]}_{1101}=R^{[1]}_{1100}=R^{[1]}_{1000}=R^{[1]}_{0100}=R^{[1]}_{0000}=1\\
&R^{[2]}_{111}=R^{[2]}_{110}=R^{[2]}_{100}=R^{[2]}_{010}=R^{[2]}_{000}=1.
\end{eqnarray}
All remaining elements are zero, which means that the corresponding projections are forbidden. The first index of the $T$ tensors is the physical index, where $\alpha = 0$ represents an unexcavated block and $\alpha = 1$ represents an excavated block. 
The virtual indices other than the physical one of $T$ are used to transfer the status of neighboring tensors. As in the illustration of the tensors shown in Fig.\ref{Fig:def_peps_2d}(c), the arrow on each index is used to distinguish the directions of the transferring status of the geometrical bonds, {\it i.e.} the indices with incoming arrows are carrying the status originated from their parent blocks, and the indices with outgoing arrows are carrying the status, which should be the same as their physical indices and will head to their child blocks. 
The constraints of the excavated ore have been reflected in our tensor network definition with very limited nonzero elements of the tensor. 
Under the constraint that if a block $(i_x,i_y)$ is excavated, so must its parent blocks $(i_x-1,i_y-1)$, $(i_x-1,i_y)$ and $(i_x-1,i_y+1)$, but an excavated block itself does not have to have child blocks.

\begin{figure}[tbh!]
\centering
    \includegraphics[width=0.9\linewidth]{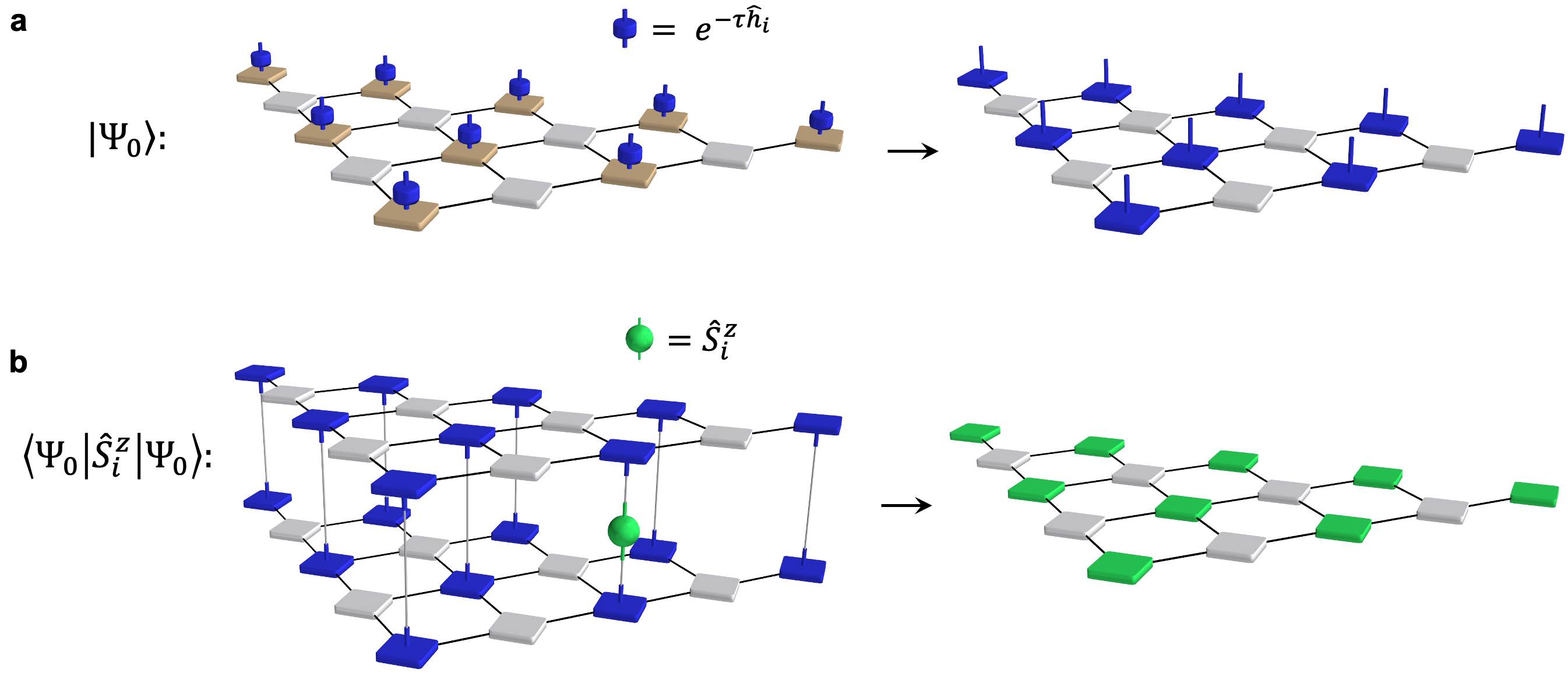}
    \caption{\textbf{The optimization process using imaginary time evolution (ITE) on the tensor network for the open-pit mining problem.} (a) The schematic of ITE process as described in Eq.(\ref{Eq:ITE1}). The left panel shows the ITE process using the local evolution gate $\exp(-\tau\hat{h}_i)$ with $\hat{h}_i = \frac{w_i}{2}(\hat{Z}_i-\hat{I})$, as shown in blue, to project the tensor network state, as shown in light brown and grey square defined in Fig.\ref{Fig:def_peps_2d}. The right panel shows the obtained ground state wavefunction, with each blue block being an updated local tensor. (b) The schematic of measuring a local operator as described in Eq.(\ref{Eq:ITE2}) or Eq.(\ref{Eq:ITE4}). The operator may change to different form. The left panel shows the tensor network construction of the expectation value calculation. The right panel shows the scalar tensor network for the local observation with renewed local tensors. }
\label{Fig:img_peps_2d}
\end{figure}

Since there is only a very limited number of nonzero elements in the initialized tensor network, there is a lot of room for optimization in terms of computational memory and time cost. In addition, the three-dimensional open-pit mine can also be defined in a similar way, just by adjusting the local tensors $T$ and the auxiliary tensors $R$ to higher-order tensors and designing the nonequivalent tensor for boundary conditions in the three-dimensional mine.

The schematic of ITE is shown in Fig.\ref{Fig:img_peps_2d}a. As the Hamiltonian for the open-pit mining problem, the same as defined in Eq.(\ref{Eq:Hamilton}), only contains single site commuting terms, we can rewrite the evolution operator for the whole system, {\it i.e.} $e^{-\tau\hat{H}}$ in Eq.(\ref{Eq:ITE1}) into the product of the local evolution operators, {\it i.e.} $e^{-\tau\hat{H}}=\Pi_i e^{-\tau\hat{h}_i}$, with $\hat{h}_i = \frac{w_i}{2}(\hat{Z}_i-\hat{I})$. Then directly set the imaginary time $\tau$ as a relatively large number, such as $\tau=10$, to achieve the ground state in the complexity of $O(n)$.

\section{Results and discussions}\label{sec:results}

\begin{figure}[tbh!]
\centering
    \includegraphics[width=0.45\linewidth]{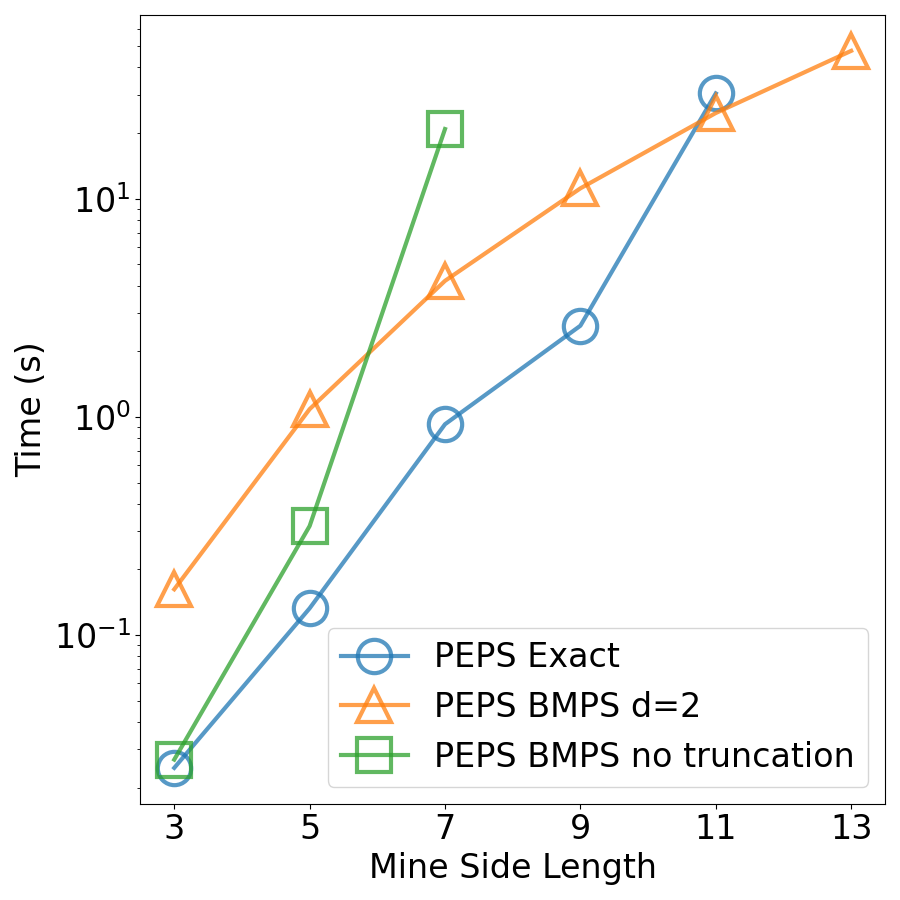}
    \caption{Computational time for two-dimensional open-pit mining problems with mine side length ranging from 3 to 13, using three different PEPS contraction approaches.}
\label{Fig:profit}
\end{figure}


 We implement our algorithm for the mining problem as an open source Python library, available at \cite{git}. Specifically, the tensor network construction described in Section \ref{sec:mining_framework} is a 2D tensor network, which can be viewed as projected entangled pair states (PEPS) \cite{VerstraeteARXIV}, with trivial physical indices on the $R$ tensors. We employ Koala \cite{pang2020efficient}, a high-performance PEPS simulation library, to perform the imaginary time evolution and calculate the expectation values by contractions.

We perform numerical experiments on open-pit mining problems with different size scales, as defined in Section \ref{sec:mining_problem}. Problem instances are generated randomly, where the value of each site follows a normal distribution with a mean of $0.1$ and a standard deviation of $1$. Each problem size is repeated five times with different mine instances to account for the possible fluctuating running condition of the computing device. In an ideal world, the runtime should be the same given a fixed problem size since the algorithm is purely deterministic. The ITE is conducted with $\tau=6$, which is large enough to achieve the ground state, but not too large to cause numerical instability, such as exceeding the maximum allowed value of the complex type in Python. Note that one can also set $\tau$ with a heuristic related to problem size to better achieve the goals. We obtain the solutions by calculating the expectation values of the Pauli Z operators following Eq.~(\ref{Eq:ITE2}). 

Fig.\ref{Fig:profit} shows a comparison of our algorithm's runtime with different contraction approaches. If no bond dimension truncation is desired, the ``PEPS Exact" method in the figure contracts a PEPS in a generally optimal order \cite{guo2019general}. The boundary matrix product state (BMPS) method has poorly scaled performance on its own, but can be executed with bond dimension truncation to significantly lower time and memory complexities \cite{pang2020efficient}.  Following the references, we can calculate the asymptotic time complexities of the three contraction approaches. While ``PEPS Exact" and ``BMPS no truncation" both take $2^{O(\sqrt{n})}$, BMPS with truncation only takes $O(d^6n)=O(n)$, where $n$ is the number of sites. Since we need to do $n$ full contractions, the total time to calculate the expectation values with truncation is $O(n^2)$. As other parts of the algorithm all run in linear time for the open-pit mining problem, the overall time complexity is $O(n^2)$. In comparison, the state-of-the-art classical algorithm Pseudoflow finds the optimal solution in $O(|V||E|\log |V|)=O(n^2\log n)$ \cite{hochbaum2008pseudoflow}. Note that although our algorithm is asymptotically faster, its constant factor is larger, and it makes approximations when truncating bond dimensions.
In Fig.\ref{Fig:profit}, we see that truncating the bond dimension to $d=2$ incurs an additional overhead but has a polynomial runtime in contrast to the exponential runtime of both exact methods, which agrees well with our theoretical analyses. 
We compare the results produced with the optimal solution given by Pseudoflow. For all problem instances that we run, given an appropriate ITE evolution time, our algorithm can always find the optimal solution with maximized profit and zero constraint violations. Thus, we conclude that truncating the bond dimension to as small as $d=2$ does not affect the accuracy of our algorithm.

It is worth noting that the open-pit mining problem has a few desirable properties for our algorithm. First, the system has a large energy gap between the ground state and the first excited state, allowing ITE to efficiently separate the optimal solution. Second, local and structured constraints decrease the complexity of constructing the tensor networks and make them reducible to well-studied tensor network forms. Third, the Hamiltonian contains only local interactions, resulting in well-controllable long-range entanglement, thus allowing aggressive bond-dimension truncations. It would be intriguing to see how the algorithm performs for more complicated and less suitable problems. Furthermore, by increasing the initial bond dimension, our algorithm can be extended beyond binary variables. A study on the performance sacrifice due to the increased initial bond dimension would be of interest. Moreover, as most of the tensor elements are zero in our construction, sparse tensor techniques could be applied, in addition to parallel computing implementations, to further improve performance.

\section{Acknowledgement}

The authors thank Mario Motta, Joseph A. Latone, Jay Borenstein, Sreeram Venkatarao, Meltem Tolunay, Allyson Stoll and Stanford CS210 class. The code used to generate the data presented in this study can be publicly accessed on GitHub at \cite{git}. This work was supported by the Department of Energy, Office of Science, Basic Energy Sciences, Materials Sciences and Engineering Division, under Contract No. DE-AC02-76SF00515.

\bibliographystyle{unsrt}
\bibliography{ref}

\end{document}